\documentclass[twocolumn,tighten]{aastex631}

\usepackage{amsmath,amstext,amssymb}
\usepackage{mathrsfs}
\usepackage{float}

\usepackage{afterpage}
\usepackage{xcolor}
\usepackage{threeparttable}

\usepackage{graphicx}
\usepackage{epsfig}

\usepackage{booktabs}
\usepackage{multirow}

\usepackage{accents}

\begin{document}

\title{Covariant Guiding Center Equations for Charged Particle Motions in General Relativistic Spacetimes}

\author{Tyler Trent}
\affiliation{School of Physics, Georgia Institute of Technology, 837 State St NW, Atlanta, GA 30332, USA}
\affiliation{Departments of Astronomy and Physics, University of Arizona, 933 N. Cherry Ave., Tucson, AZ 85721, USA}

\email{ttrent@arizona.edu}

\author{Karin Roley}
\affiliation{School of Physics, Georgia Institute of Technology, 837 State St NW, Atlanta, GA 30332, USA}

\author{Matthew Golden}
\affiliation{School of Physics, Georgia Institute of Technology, 837 State St NW, Atlanta, GA 30332, USA}

\author{Dimitrios Psaltis}
\affiliation{School of Physics, Georgia Institute of Technology, 837 State St NW, Atlanta, GA 30332, USA}

\author{Feryal \"Ozel}
\affiliation{School of Physics, Georgia Institute of Technology, 837 State St NW, Atlanta, GA 30332, USA}

\begin{abstract}
Low density plasmas in curved spacetimes, such as those found in accretion flows around black holes, are challenging to model from first principles, owing to the large scale separation between the characteristic scales of the microscopic processes and large mean-free-paths comparable to the system sizes. Kinetic approaches become necessary to capture the relevant physics but lack the dynamic range to model the global characteristics of the systems. In this paper, we develop new covariant guiding center equations of motion for charges in general relativistic spacetimes that are computationally tractable. We decompose the particle motion into a fast gyration, which we integrate analytically and a slow drift of the guiding center, which can be solved numerically. We derive covariant conservation laws for the motions of the guiding centers and show, through a number of limiting cases, that the equations contain all known drift mechanisms. Finally, we present the general relativistic expressions for the various drift velocities in Schwarzschild spacetimes.
\end{abstract}

\section{Introduction}

Rarefied plasmas are found in a multitude of astrophysical environments ranging from the heliosphere to the intracluster medium and the accretion flows around black holes. They are often modeled as fluids using equations of Magnetohydrodynamics (MHD) or, as needed, of General Relativistic magnetohydrodynamics (GRMHD). The computational efficiency of a fluid approach allows for global simulations and has proven useful for understanding the overall dynamics of the plasma in these systems. As one particular example, GRMHD simulations have guided the Event Horizon Telescope (EHT) imaging observations of the black hole at the center of the M87 galaxy and Sagittarius A* (SgrA*) and enabled the reconstruction and interpretation of these images \citep{EHT_M87_1,EHT_SGRA_1,Medeiros_2023}. 

Despite their usefulness, fluid approaches are not formally valid in the weakly collisional (or collisionless) conditions of these rarefied plasmas. The long mean-free-paths of charged particles along field lines are expected to lead to large pressure anisotropies and introduce non-local couplings of the dynamics and thermodynamics of the plasmas. This shortcoming of fluid approaches is especially important when modeling, e.g., the observable effects of episodic magnetic reconnection events that accelerate electrons to ultra-relativistic non-thermal energies and are believed to be responsible for the X-ray flares observed from Sgr~A*. Such events have been traditionally modeled by injecting locally a distribution of ultra-relativistic electrons into the background of a GRMHD simulation and then calculating their radiative properties without accounting for their dynamics~\citep[see, e.g.,][]{Dodds-Eden_2010,Ball_2016,Chatterjee_2021,Scepi_2022}. This local approach ignores the mean-free paths of the ultra-relativistic particles along magnetic field lines, which can be significantly larger than the macroscopic (i.e., horizon) scales of the systems. Perhaps more importantly, it does not address the apparent tension with imaging observations of the flare emission that appears to require these ultrarelativistic particles to remain trapped in compact regions and orbit around the black hole~\citep{GRAVITY}.

To overcome these shortcomings, kinetic approaches, such as Particle-in-Cell (PIC), describe collisionless plasmas from first principles by self-consistently solving for the individual motions of charged particles and the fields they produce. These approaches are capable of resolving most of the microphysics that fluid approaches cannot capture, such as the acceleration of charged particles in magnetic reconnection events \citep[see, e.g.,][]{Sironi_2014,Werner2018,Sironi_2020} and instabilities that form from pressure anisotropies \citep{Kunz_2014, Zhdankin_2023}. Hybrid particle-MHD methods have also been developed to model the dynamics of relativistic particles within MHD backgrounds~\citep[see, e.g.,][]{Bai2015,Mignone2018,vanMarle2018}. In several situations of astrophysical interest, however, the large scale separation between the kinetic scales, i.e., the radius of gyration around local magnetic fields (gyroradius) and the plasma skin depth, and the macroscopic length scales of astrophysical systems, such as the size of the accretion flow, constrains the dynamic range that can be simulated numerically. For this reason, current state-of-the-art kinetic and hybrid simulations of plasmas around compact objects have largely been limited to local simulations that study, e.g., magnetic reconnection rates and the particles they accelerate \citep{Werner2018,Ball_2018, Ball_2019, Sironi_2021, Hakobyan_2023}. At the opposite extreme, global PIC simulations of black hole accretion disks use nonphysical parameters to overcome the scale separation \citep{Parfrey_2019, Crinquand_2020, Crinquand_2021, Crinquand_2022, Galishnikova_2023}.

Fortunately, the scale separation in the gyroradius and macroscopic length scales puts these systems in a regime where the kinetic approach can be employed in the the guiding center approximation. The guiding center formalism decomposes the motion of charged particles in strong magnetic fields into a fast gyration (sometimes referred to as gyromotion) about the local magnetic field lines and the slow drift of the guiding center (the center of the gyromotion, see e.g., \citealt{Northrop_book}). This is justified in a strong and slowly varying magnetic field, where the particle undergoes nearly uniform circular motion with the center of that circle slowly drifting. Earlier work analytically solved for the guiding center velocities (sometimes referred to as the drift velocity) in flat spacetimes, which allows for studies of just the drift motion of the particle without including the trivial and uninformative gyromotion~\citep[see][for a review]{Cary2009}. Causes of drifts perpendicular to the magnetic field lines include the presence of an electric field $\vec{E}$ (the $\vec{E}\times\vec{B}$ drift), the presence of a gravitational field (the $\vec{g}\times\vec{B}$ drift), and a gradient in the magnetic field strength (the $\vec{\nabla} B\times \vec{B}$ drift). This approach has been successfully implemented in a number of simulations where test particle trajectories are calculated using the guiding center approximation with a MHD background (see, e.g, \citealt{Gordovskyy_2014, Threlfall_2015, Ripperda_2017, Borissov_2020, Gordovskyy_2023,Mignone2023}). The length scale in which the guiding center evolves is much greater than the gyromotion, allowing for a coarser numerical resolution and a significant numerical speed up in the calculation of the particle motion. This in turn, enables tracking of these particles throughout the global system, which otherwise would not have been computationally possible.

To date, there has not been a general relativistic covariant formulation of the guiding center approach. One attempt by \cite{Beklemishev_2004} uses the tetrad formalism and leaves the equations in a form that has not been computationally tractable. In another drift-kinetic study in general relativistic spacetimes \citep{Bacchini_2020}, only the $\Vec{E}\times\Vec{B}$ drift velocity was incorporated by using the special-relativistic expression to evaluate it in a local Lorentz frame and then boosting it to the coordinate frame. This method does not allow for magnetic mirroring of charged particles and is not easily generalizable to include drifts that cannot be readily written in terms of quantities evaluated in a local Lorentz frame, such as those arising from nonuniform electromagnetic fields or from gravity (which vanishes in the local Lorentz frame). 

\citet{Trent2023} introduced a set of covariant, general relativistic guiding center equations that can be used in arbitrary spacetimes and for arbitrary particle velocities. The key insight in that treatment came from the realization that it is possible to circumvent the difficulty of obtaining covariant equations for the various drift {\em velocities} by deriving, instead, a covariant {\em acceleration\/} equation for the motion of the gyrocenter. We then used a set of numerical experiments in different configurations to demonstrate that the new covariant formalism incorporates all of the drifts described by flat-spacetime guiding-center equations. 

In this paper, we present a detailed analytic exploration of the new covariant guiding-center equations for charged particles in arbitrary spacetimes. In \S \ref{sec:derivation}, we outline the detailed derivation of the acceleration equation for the guiding center in the presence of an arbitrary electromagnetic field. In \S \ref{sec: conservation laws}, we derive an integral of motion for the guiding center and show that it is an analog to the unity norm of the 4-velocity of a relativistic particle. In \S \ref{sec: limiting cases}, we consider analytically a number of limiting cases and show that the guiding center equations recover the standard drift velocities. We then generalize the expressions for the drift velocities to the case of motion in a Schwarzschild spacetime. Finally, in \S \ref{sec:conclusion}, we summarize our results and discuss possible applications of these general relativistic guiding center equations.

\section{Covariant Guiding Center Equations}
\label{sec:derivation}
In this section we derive the covariant guiding center equation of motion. We start from the equation of motion for a charged particle in a general spacetime with an arbitrary electromagnetic field
\begin{align}
    \label{eqn:full EOM}
    \frac{d^2x^\alpha}{d\tau^2} = -\Gamma^\alpha_{\mu\nu} \frac{dx^\mu}{d\tau} \frac{dx^\nu}{d\tau} +\frac{q}{m}F^{\alpha}_{~ \beta}\frac{d x^\beta}{d\tau},
\end{align}
where $x^\alpha$ is the four-position of the particle, $\tau$ is the proper time, $\Gamma^\alpha_{\mu\nu}$ are the Christoffel symbols, $q/m$ is the charge-to-mass ratio of the particle, and $F^\alpha_{~\beta}$ is the electromagnetic field tensor. Throughout this section, we use units such that $G=c=1$.

For systems where the electromagnetic field is slowly varying in time and space and the effects due to the curvature of spacetime are much weaker than the effects from the electromagnetic force, the particle undergoes a nearly uniform circular motion with a slow drift. We refer to this circular motion as the gyromotion and to the motion of the guiding center as the drift. The goal of the guiding center approximation is to decompose the motion of the particle perpendicular to the magnetic field into a fast gyromotion and a slow drift of the guiding center. This allows us to integrate out the gyromotion analytically and formulate an equation of motion only for the guiding center.

We can carry out this approximation under the following conditions, (extended from \citealt{Vandervoort_1960}):\\
\noindent 1. The gyroradius $\rho$ is significantly smaller than the characteristic scale over which the electromagnetic field varies
\begin{equation}
\label{eqn:assumption 1}
    \rho\ll|F^\alpha_{~\beta}| / \left|\frac{\partial F^\alpha_{~\beta}}{\partial x^\mu}\right|\;.
\end{equation}
\noindent 2. The particle drifts for many gyroperiods before the field changes considerably,
\begin{equation}
\label{eqn:assumption 2}
    \frac{1}{\omega}\left|\frac{\partial\chi^\nu}{\partial \tau}\right|\ll|F^\alpha_{~\beta}| / \left|\frac{\partial F^\alpha_{~\beta}}{\partial x^\mu}\right|\;,
\end{equation}
where $\omega$ is the gryofrequency and $d\chi/d\tau$ is the guiding center four-velocity. \\
\noindent 3. The effect of the spacetime curvature on the motion of the particle is weaker than that of the electromagnetic field,
\begin{equation}
\label{eqn:assumption 3}
    \left|\Gamma^\alpha_{\mu\nu}\frac{dx^\mu}{d\tau} \frac{dx^\nu}{d\tau}\right|\ll \frac{q}{m} \left|F^\alpha_{~\beta}\frac{dx^\beta}{d\tau}\right|\;.
\end{equation}
In the above expressions, we use the double vertical bar symbol $|~|$ to denote the magnitude of a typical component. The above assumptions require that the derivatives of the electromagnetic field along the path of the guiding center and the Christoffel symbols are all considered first order perturbations to the guiding center motion compared to the effects of the electromagnetic field tensor, which are of zeroth order.

Because the final equation of motion will only describe the guiding center, we evaluate all quantities at the position of the guiding center. The distance between the position of the particle and that of the guiding center is equal to the gyroradius. We expand the electromagnetic field tensor around the guiding center to first order along coordinate lines, which involves only a partial derivative. However, because we assume that the gravity term is much smaller than the Lorentz force term, we evaluate the Christoffel symbols at the guiding center. This amounts to a zeroth-order expansion of the Christoffel symbols, given that the gravity term is already a first order term in our approximation. The resulting equation becomes
\begin{align}
\label{eqn:expansion}
    \underbrace{\frac{d^2x^\alpha}{d\tau^2}}_{\mathcal{O}(0)}=
    \underbrace{\left.-\Gamma^\alpha_{\mu\nu}\right\vert_{\chi} \frac{dx^\mu}{d\tau} \frac{dx^\nu}{d\tau}}_{\mathcal{O}(1)}
    +\underbrace{\frac{q}{m}\left.F^{\alpha}_{~ \beta}\right\vert_{\chi}\frac{d x^\beta}{d\tau}}_{\mathcal{O}(0)}\nonumber\\
    +\underbrace{\frac{q}{m}\left.\frac{\partial F^{\alpha}_{~ \beta}}{\partial x^\mu}\right\vert_{\chi}\left(x^\mu-\chi^\mu\right)\frac{d x^\beta}{d\tau}}_{\mathcal{O}(1)}.
\end{align}

To derive the guiding center equation of motion, we take an iterative approach by first solving the zeroth order system, which describes the motion of a charged particle in a constant electromagnetic field and spacetime. The solution to the zeroth order system then gives us the necessary building blocks to construct our ansatz for the first order system. The first order system includes a general spacetime with arbitrary electromagnetic field that are consistent with equations \eqref{eqn:assumption 1}-\eqref{eqn:assumption 3}. Ultimately, we solve for the acceleration of the guiding center.

\subsection{Zeroth Order Approximation}

To zeroth order, the equation of motion~(\ref{eqn:expansion}) simplifies to that of a charged particle in a constant spacetime metric with a constant electromagnetic field, i.e.,
\begin{equation}
    \label{eqn:EOM flatspacetime}
    \frac{d^2x^\alpha}{d\tau^2} = \frac{q}{m} F^{\alpha}_{~\beta}\frac{d x^\beta}{d\tau}.
\end{equation}
Since this is a homogeneous equation, the most convenient way to obtain an analytic solution is by using the eigenvalues and eigenvectors of the tensor $(q/m)\,F^\alpha_{~\beta}$ \citep{Vandervoort_1960}. In the next subsection, we introduce the eigensystem of the electromagnetic field tensor before deriving the solution to the differential equation.

\subsubsection{Eigensystem of the Electromagnetic Field Tensor}

We denote the eigenvectors of $q/m\,F^{\alpha}_{~\beta}$ by $\{\sigma^\beta,\delta^\beta,\psi^\beta,\Upsilon^\beta\}$ and their eigenvalues by $\{i\omega,-i\omega,\lambda,-\lambda\}$, respectively. Here, $\omega$ and $\lambda$ are real numbers, the eigenvectors $\sigma$ and $\delta$ are complex, and the eigenvectors $\psi$ and $\Upsilon$ are real. We conveniently write the eigenvalues in terms of the field invariants, as \citep{Fradkin_1978}
\begin{align}
\label{eqn:eignevalues}
    \omega  &= \frac{q}{2m}\biggl\{F^{\alpha\beta}F_{\alpha\beta}+\Bigl[(F^{\alpha\beta}F_{\alpha\beta})^2+(F^{\alpha\beta}F^D_{\alpha\beta})^2\Bigr]^{1/2}\biggr\}^{1/2},\\
    \lambda &= \frac{q}{2m}\biggl\{-F^{\alpha\beta}F_{\alpha\beta}+\Bigl[(F^{\alpha\beta}F_{\alpha\beta})^2+(F^{\alpha\beta}F^D_{\alpha\beta})^2\Bigr]^{1/2}\biggr\}^{1/2},
\end{align}
where $F^D_{\alpha\beta}$ stands for the dual of $F_{\alpha\beta}$. 

We construct the eigenvectors by invoking the Cayley-Hamilton Theorem, which shows that every square matrix satisfies its own characteristic polynomial, i.e., that
\begin{align}
    \label{eqn:Cayley-Hamilton Theorem}
    (F^{\alpha}_{~\beta}-i\omega I^{\alpha}_{\beta})(F^{\beta}_{~\zeta}+i\omega I^{\beta}_{\zeta})(F^{\zeta}_{~\mu}-\lambda I^{\zeta}_{\mu})(F^{\mu}_{~\nu}+\lambda I^{\mu}_{\nu})=0,
\end{align}
where $I$ is the four-by-four identity matrix. The non-normalized eigenvector for a given eigenvalue is then the subset of the product of the three binomials from the characteristic polynomial that correspond to the other eigenvalues. We write the non-normalized eigenvectors as
\begin{align}
    \label{eqn:eigenvectors}
    \Tilde{\sigma}^\alpha =& (F^{\alpha}_{~\beta}+i\omega I^{\alpha}_{\beta})(F^{\beta}_{~\mu}-\lambda I^{\beta}_{\mu})(F^{\mu}_{~\nu}+\lambda I^{\mu}_{\nu})e^\nu,\\
    \Tilde{\delta}^\alpha =& (F^{\alpha}_{~\beta}-i\omega I^{\alpha}_{\beta})(F^{\beta}_{~\mu}-\lambda I^{\beta}_{\mu})(F^{\mu}_{~\nu}+\lambda I^{\mu}_{\nu})e^\nu,\\ 
    \Tilde{\psi}^\alpha =& (F^{\alpha}_{~\beta}-i\omega I^{\alpha}_{\beta})(F^{\beta}_{~\mu}+i\omega I^{\beta}_{\mu})(F^{\mu}_{~\nu}+\lambda I^{\mu}_{\nu})e^\nu,\\
    \label{eqn:eigenvectors end}    
    \Tilde{\Upsilon}^\alpha =& (F^{\alpha}_{~\beta}-i\omega I^{\alpha}_{\beta})(F^{\beta}_{~\mu}+i\omega I^{\beta}_{\mu})(F^{\mu}_{~\nu}-\lambda I^{\mu}_{\nu})e^\nu,
\end{align}
where $e^\nu$ is any non-zero four-vector.

When there are no degenerate eigenvalues, we normalize the eigenvectors such that $\sigma^\alpha \delta_\alpha = 1$ and $\psi^\alpha\Upsilon_\alpha=1$. When an eigenvalue is degenerate, we normalize the corresponding eigenvector to 1 by contracting it with itself instead.

\subsubsection{Zeroth-order Solution of Charged Particle Trajectory}

\label{sec:flat spacetime solution}
To obtain a solution to equation \eqref{eqn:EOM flatspacetime}, we start by projecting the four-velocity onto the normalized eigenbasis\footnote{If there are degenerate eigenvalues, given the symmetry of the eigensystem, the eigenvectors can always be constructed to be linearly independent.}. We then insert the expanded four-velocity into equation~\eqref{eqn:EOM flatspacetime} and obtain
\begin{align}
    \label{eqn:expanding flat EOM1}
     \frac{d}{d\tau}\biggl(
     \frac{dx^\mu}{d\tau} \delta_\mu \sigma^\alpha
     +\frac{dx^\mu}{d\tau} \sigma_\mu \delta^\alpha
     +\frac{dx^\mu}{d\tau} \Upsilon_\mu \psi^\alpha
    +\frac{dx^\mu}{d\tau} \psi_\mu \Upsilon^\alpha
     \biggr)=\nonumber\\
    F^{\alpha}_{~\beta}\biggl(
     \frac{dx^\mu}{d\tau} \delta_\mu \sigma^\beta
     +\frac{dx^\mu}{d\tau} \sigma_\mu \delta^\beta
     +\frac{dx^\mu}{d\tau} \Upsilon_\mu \psi^\beta
    +\frac{dx^\mu}{d\tau} \psi_\mu \Upsilon^\beta
     \biggr).
\end{align}
Note that we perform the expansion onto the eigenbasis in the above equation because the eigenvectors are null vectors.

On the right-hand side, we contract the eigenvectors with the field tensor to obtain the eigenvalues. On the left-hand-side, we move the eigenvectors outside of the derivative, since the field tensor is constant to zeroth order, and obtain
\begin{widetext}
\begin{align}
     \label{eqn:expanding flat EOM2}
     \frac{d}{d\tau}\biggl(\frac{dx^\mu}{d\tau} \delta_\mu \biggr)\sigma^\alpha
     +\frac{d}{d\tau}\biggl(\frac{dx^\mu}{d\tau} \sigma_\mu \biggr)\delta^\alpha
     &+\frac{d}{d\tau}\biggl(\frac{dx^\mu}{d\tau} \Upsilon_\mu \biggr)\psi^\alpha
    +\frac{d}{d\tau}\biggl(\frac{dx^\mu}{d\tau} \psi_\mu\biggr) \Upsilon^\alpha
    \\ \nonumber
     &=i\omega\biggl(\frac{dx^\mu}{d\tau} \delta_\mu\biggr) \sigma^\alpha
     \,-\,i\omega\biggl(\frac{dx^\mu}{d\tau} \sigma_\mu\biggr) \delta^\alpha
     ~+\,\lambda\,\biggl(\frac{dx^\mu}{d\tau} \Upsilon_\mu\biggr) \psi^\alpha
     ~-\,\lambda\,\biggl(\frac{dx^\mu}{d\tau} \psi_\mu \biggr)\Upsilon^\alpha.
\end{align}
\end{widetext}

Using the linear independence of the eigenvectors, we turn this equation into four separate differential equations, each of which has an exponential solution. The general solution is the sum of all four, i.e.,
\begin{widetext}
\begin{align}
\label{eqn:velocity solution}
    \frac{dx^\alpha}{d\tau}(\tau) = i\omega\rho_0e^{i\omega\tau}\sigma^\alpha
    -i\omega\rho_0^*e^{-i\omega\tau}\delta^\alpha
    &+\Upsilon_\beta\frac{dx^\beta}{d\tau}\bigg|_{\tau=0}e^{\lambda\tau} \psi^\alpha
    +\psi_\beta\frac{dx^\beta}{d\tau}\bigg|_{\tau=0}e^{-\lambda\tau} \Upsilon^\alpha,\\
    \text{Integrating this once over proper time, we obtain}\nonumber\\
\label{eqn:position solution}
    x^\alpha(\tau) = \underbrace{\rho_0e^{i\omega\tau}\sigma^\alpha
    +\rho_0^*e^{-i\omega\tau}\delta^\alpha}_\text{gyromotion}
    &+\underbrace{ \frac{1}{\lambda}\Upsilon_\beta\frac{dx^\beta}{d\tau}\bigg|_{\tau=0}e^{\lambda\tau} \psi^\alpha
    -\frac{1}{\lambda} \psi_\beta\frac{dx^\beta}{d\tau}\bigg|_{\tau=0}e^{-\lambda\tau} \Upsilon^\alpha
    +\kappa^\alpha}_\text{drift}.
\end{align}
\end{widetext}
Here, $\kappa^\alpha$ is a constant determined by the initial conditions and we defined the constant
\begin{equation}
\label{eqn:gyroradius}
    \rho_0\equiv \biggl(\frac{i\sigma_\beta}{\omega}\frac{dx^\beta}{d\tau}\biggr)\biggl\vert_{\tau=0}\;.
\end{equation}

We can interpret the solution in equation \eqref{eqn:position solution} as consisting of two parts: a gyromotion and a drift motion. The oscillatory terms represent the gyromotion, with $\omega$ as the gyrofrequency and $\sigma$ and $\delta$ as four-vectors that live in the gyromotion plane. The complex quantity $\rho_0$ defined above has a magnitude that is equal to the gyroradius of motion divided by $\sqrt{2}$ (because the gyromotion is decomposed into two orthogonal gyromotions along the eigenvectors $\sigma^\alpha$ and $\delta^\alpha$) and a phase that corresponds to the initial azimuthal position of the particle on the circle of gyration. The exponential terms represent the drift (or guiding-center) motion at zeroth order. 

\subsection{First Order Approximation}
\label{subsection: generalized GCA}
When the requirements in equations~\eqref{eqn:assumption 1}-\eqref{eqn:assumption 3} are satisfied, we can decompose the first-order solution to equation \eqref{eqn:full EOM} into a gyromotion component, which remains of zeroth order, and a drifting guiding center component, which will be of first order. Retaining only the gyromotion terms from the zeroth-order equation \eqref{eqn:position solution} and allowing for the drift term to be arbitrary, we construct the following ansatz
\begin{align}
    \label{eqn:four-position definition}
    x^\alpha(\tau) = \rho(\tau) \sigma^\alpha
    +\rho^*(\tau)\delta^\alpha
    +\chi^\alpha\;,
\end{align}
where $\chi^\alpha$ is the four-position of the guiding center. Here, $\rho(\tau)$ is the instantaneous complex gyroradius. Its magnitude is determined by the conservation of the magnetic moment (see \S\ref{sec: conservation laws})
\begin{equation}
\mu\equiv \rho_0^2 \omega_0 = \rho^2 \omega   
\end{equation}
such that
\begin{equation}
\label{eq:rho_tau}
    \rho(\tau)=\rho_0\sqrt{\frac{\omega_0}{\omega}}e^{-i\omega\tau}\;.
\end{equation} 
Taking the time derivative of the ansatz gives the four-velocity
\begin{align}
\label{eqn:four-velocity full}
    \frac{dx^\alpha}{d\tau}=& \,i(\omega+\frac{d\omega}{d\tau}\tau)\rho\sigma^\alpha +i\omega\rho\frac{d\sigma^\alpha}{d\tau}\\ \nonumber
    &-i(\omega+\frac{d\omega}{d\tau}\tau)\rho^*\delta^\alpha-i\omega\rho^*\frac{d\delta^\alpha}{d\tau} + \frac{d\chi}{d\tau}.
\end{align}
Our second assumption, equation \eqref{eqn:assumption 2}, implies that derivatives of the electromagnetic field along the path of the particle are small. This is equivalent to assuming that total derivatives with respect to proper time of the eigenvalues and eigenvectors are of first order in our approximation. Dropping all terms in equation~(\ref{eqn:four-velocity full}) that are of second order, we obtain
\begin{align}
\label{eqn:four-velocity}
    \frac{dx^\alpha}{d\tau}\approx \,i\omega\rho\sigma^\alpha
    -i\omega\rho^*\delta^\alpha+ \frac{d\chi^\alpha}{d\tau}.
\end{align}
Similarly, we find the four-acceleration to first order as
\begin{align}
    \label{eqn:four-acceleration}
    \frac{d^2x^\alpha}{d\tau^2}\approx \, -\omega^2\rho\sigma^\alpha
    -\omega^2\rho^*\delta^\alpha + \frac{d^2\chi^\alpha}{d\tau^2}.
\end{align}

Inserting equation~(\ref{eqn:four-position definition}) for the position of the particle, equation~(\ref{eqn:four-velocity}) for the four-velocity, and equation~(\ref{eqn:four-acceleration}) for the four-acceleration into equation \eqref{eqn:expansion}, we find
\begin{widetext}
\begin{align}
\label{eqn:midway}
    -\omega^2\rho\sigma^\alpha -\omega^2\rho^*\delta^\alpha + \frac{d^2\chi^\alpha}{d\tau^2} =& 
    -\Gamma^\alpha_{\mu\nu}
    \biggl(i\omega\rho\sigma^\mu-i\omega\rho^*\delta^\mu+ \frac{d\chi^\mu}{d\tau}\biggr)
    \biggl(i\omega\rho\sigma^\nu -i\omega\rho^*\delta^\nu+ \frac{d\chi^\nu}{d\tau}\biggr) \\ \nonumber
    &\quad+\frac{q}{m}F^{\alpha}_{~ \beta}\biggl(i\omega\rho\sigma^\beta -i\omega\rho^*\delta^\beta+ \frac{d\chi^\beta}{d\tau}\biggr)
    +\frac{q}{m}\frac{\partial F^{\alpha}_{~ \beta}}{\partial x^\mu}
    \left(\rho\sigma^\mu+\rho^*\delta^\mu\right)\biggl(i\omega\rho\sigma^\beta -i\omega\rho^*\delta^\beta+ \frac{d\chi^\beta}{d\tau}\biggr).
\end{align}   
\end{widetext}

We integrate equation \eqref{eqn:midway} over one gyroperiod, which removes all oscillatory terms. As a result, all terms that are linear in $\rho$ and $\rho^*$ or contain the products $\rho\rho$ and $\rho^*\rho^*$ vanish, leaving only terms with $\rho\rho^*$ (since $e^{i\omega\tau}e^{-i\omega\tau}=1$) or $d\chi/d\tau$. The resulting equation is
\begin{align}
\label{eqn: GCA EOM unsimplified}
    \frac{d^2\chi^\alpha}{d\tau^2} =& -\Gamma^\alpha_{\mu\nu}
    \biggl( \omega\omega_0\rho_0^2\sigma^\mu\delta^\nu + \omega\omega_0\rho_0^2\sigma^\nu\delta^\mu + \frac{d\chi^\mu}{d\tau}\frac{d\chi^\nu}{d\tau}\biggr)\nonumber\\
    &+\frac{q}{m}F^{\alpha}_{~ \beta}\frac{d\chi^\beta}{d\tau}
    + i\omega_0\rho_0^2\frac{q}{m}\frac{\partial F^{\alpha}_{~ \beta}}{\partial x^\mu}(\sigma^\beta\delta^\mu-\sigma^\mu\delta^\beta).
\end{align}

We can now combine the first two terms and the last term of equation \eqref{eqn: GCA EOM unsimplified} to form a covariant derivative of the field tensor. To show this, we start with the expression for the covariant derivative of the electromagnetic field tensor
\begin{align}
    &\frac{q}{m}\nabla_\mu F^{\alpha}_{~ \beta}(\sigma^\beta\delta^\mu-\sigma^\mu\delta^\beta) =\nonumber\\ 
    &\frac{q}{m}\frac{\partial F^{\alpha}_{~ \beta}}{\partial x^\mu}(\sigma^\beta\delta^\mu-\sigma^\mu\delta^\beta)
    +\frac{q}{m}\Gamma^{\alpha}_{\mu\nu}F^\nu_{~\beta}
    (\sigma^\beta\delta^\mu-\sigma^\mu\delta^\beta).
\end{align}
In the last term, we contract the $\beta$ index of the eigenvectors with the field tensor to get the corresponding eigenvalues
\begin{align}
\label{eq:F cov der}
    \frac{q}{m}\nabla_\mu F^{\alpha}_{~ \beta}(\sigma^\beta\delta^\mu-\sigma^\mu\delta^\beta) =&
    \frac{q}{m}\frac{\partial F^{\alpha}_{~ \beta}}{\partial x^\mu}(\sigma^\beta\delta^\mu-\sigma^\mu\delta^\beta)\quad\nonumber\\ 
    &+i\omega\Gamma^{\alpha}_{\mu\nu}
    (\sigma^\nu\delta^\mu +\sigma^\mu\delta^\nu).
\end{align}
Comparing equations~(\ref{eqn: GCA EOM unsimplified}) and (\ref{eq:F cov der}) allows us to write the equation of motion for the guiding center in a manifestly covariant form
\begin{align}
    \label{eqn:GCA EOM covariant}
    \frac{d^2\chi^\alpha}{d\tau^2} =& -\Gamma^\alpha_{\mu\nu}
    \frac{d\chi^\mu}{d\tau}\frac{d\chi^\nu}{d\tau}
    +\frac{q}{m}F^{\alpha}_{~ \beta}\frac{d\chi^\beta}{d\tau}\nonumber\\
    &+ i\omega_0\rho_0^2\frac{q}{m} \nabla_\mu F^{\alpha}_{~ \beta}(\sigma^\beta\delta^\mu-\sigma^\mu\delta^\beta).
\end{align}

Equation~(\ref{eqn:GCA EOM covariant}) is the equation of motion presented in \citet{Trent2023}. Even though the last term has an imaginary coefficient, the complex nature of the eigenvectors render it real. We can simplify it further and put it in a form that is manifestly real and will allow us in \S\ref{sec: conservation laws} to derive a conservation law by using the homogeneous Maxwell equation
\begin{equation}
    \nabla_{\mu}F_{\alpha\beta} + \nabla_{\alpha}F_{\beta\mu} + \nabla_{\beta}F_{\mu\alpha} = 0\;.
\end{equation}
Substituting this into the last term of equation~\eqref{eqn:GCA EOM covariant}, we obtain
\begin{align}
    &(- \frac{q}{m} \nabla_{\alpha}F_{\beta\mu} 
    -  \frac{q}{m} \nabla_{\beta}F_{\mu\alpha})
    (\sigma^\beta\delta^\mu
    -\sigma^\mu\delta^\beta)\nonumber\\
    &\qquad= 
    \frac{q}{m}\nabla_{\alpha}F_{\mu\beta}\sigma^\beta\delta^\mu.
\end{align}

The $\alpha$ index on the covariant derivative allows us to fully contract the field tensor with the eigenvectors by using the product rule:
\begin{align}  
    \frac{q}{m}\nabla^{\alpha}F^\mu_{~\beta}\sigma^\beta\delta_\mu \nonumber 
    &= \frac{q}{m}\nabla^{\alpha}(F^\mu_{~\beta}\sigma^\beta)\delta_\mu
    -\frac{q}{m}F^\mu_{~\beta}\nabla^{\alpha}(\sigma^\beta)\delta_\mu\nonumber\\
    &=\nabla^{\alpha}(i\omega\sigma^\mu)\delta_\mu
    -i\omega\delta^\beta \nabla^{\alpha}(\sigma_\beta)\nonumber\\
    \label{eqn:derivative of eigenvalue}
    &= i\nabla^{\alpha}\omega.
\end{align}
Finally, substituting equation~\eqref{eqn:derivative of eigenvalue} into equation~\eqref{eqn:GCA EOM covariant}, we obtain
\begin{align}
\label{eqn:GCA EOM Final}
    \frac{d^2\chi^\alpha}{d\tau^2} =& -\Gamma^\alpha_{\mu\nu}
    \frac{d\chi^\mu}{d\tau}\frac{d\chi^\nu}{d\tau}
    +\frac{q}{m}F^{\alpha}_{~ \beta}\frac{d\chi^\beta}{d\tau}
    - \omega_0\rho_0^2\nabla^{\alpha}\omega.
\end{align}
This equation shows that, to first order, the covariant guiding center equation of motion differs from the one that describe the full motion of the particle only by a single term involving the derivative of the scalar gyrofrequency.

\hfill
\section{Conservation Laws}
\label{sec: conservation laws}

In flat spacetime, integrating the gyromotion of a charged particle results in the adiabatic conservation of the  magnetic moment, usually defined as $\mu=v_\perp^2/2B$, where $v_\perp$ is the velocity of the charge in the direction perpendicular to the magnetic field $B$. \cite{Fradkin_1978} showed that a covariant generalization of the definition of the magnetic moment can be derived as
\begin{equation}
    \label{eq:magnetic moment}
    \mu=\frac{1}{2 m\omega} F^\alpha_{\;\beta} {\cal A}^\beta_{\;\alpha}\;,
\end{equation}
where $m$ is the mass of the charge, $\omega$ is the appropriate eigenvalue of $(q/m)F^\alpha_{\;\beta}$, as before, and the antisymmetric electromagnetic moment is defined as
\begin{equation}
\label{eq:moment tensor}
{\cal A}^{\alpha\beta}\equiv \frac{1}{2}\frac{q}{c}
\left(x^\alpha \frac{dx^\beta}{d\tau}-x^\beta \frac{dx^\alpha}{d\tau}\right)\;.
\end{equation}
In this expression, the position vector $x^\alpha$ is understood to include only the gyration and not the motion of the guiding center. 

Writing, in general, the equation of the gyromotion as 
\begin{equation}
x^\alpha=\hat{\rho}(\tau)e^{i\omega\tau}\sigma^\alpha+\hat{\rho}^*(\tau)e^{-i\omega\tau}\delta^\alpha
\end{equation}
and inserting this expression into equations~(\ref{eq:magnetic moment})-(\ref{eq:moment tensor}), we obtain $\mu=\hat{\rho}^2\omega$. This is the origin of equation~(\ref{eq:rho_tau}) that we used before.

The guiding center is a virtual position of a particle and, therefore, does not have to obey any of the conservation laws that apply to actual particles. The most trivial conservation law for massive particles is the norm of the four-velocity, $u^\alpha u_\alpha=-1$. Interestingly, the guiding center has an analog to this conservation law, as we derive in this section.

We contract the acceleration equation of the guiding center velocity, equation \eqref{eqn:GCA EOM Final}, with the four-velocity of the guiding center to obtain
\begin{align}
    \frac{d\chi_\alpha}{d\tau}\frac{d^2\chi^\alpha}{d\tau^2} =& -\frac{d\chi_\alpha}{d\tau}\Gamma^\alpha_{\mu\nu}
    \frac{d\chi^\mu}{d\tau}\frac{d\chi^\nu}{d\tau}\nonumber\\
    &\quad+\frac{d\chi_\alpha}{d\tau}\frac{q}{m}F^{\alpha}_{~ \beta}\frac{d\chi^\beta}{d\tau}
    - \frac{d\chi_\alpha}{d\tau}\omega_0\rho_0^2\nabla^{\alpha}\omega.
\end{align}

The term containing the field tensor is zero due to its anti-symmetry. We then simplify the left hand side using the product rule,
\begin{align}
    \frac{1}{2}\frac{d}{d\tau}\biggl(\frac{d\chi^\alpha}{d\tau}\frac{d\chi_\alpha}{d\tau}\biggr) =& -\frac{d\chi_\alpha}{d\tau}\Gamma^\alpha_{\mu\nu}
    \frac{d\chi^\mu}{d\tau}\frac{d\chi^\nu}{d\tau}
    - \frac{d\chi_\alpha}{d\tau}\omega_0\rho_0^2\nabla^{\alpha}\omega,
\end{align}
and combine the left hand side and the term with the Christoffel symbols to form the directional covariant derivative such that
\begin{align}
    \frac{d\chi^\alpha}{d\tau}\nabla_{\alpha}\biggl(\frac{d\chi^\beta}{d\tau}\frac{d\chi_\beta}{d\tau} + 2\omega_0\rho_0^2\omega \biggr) =0.
\end{align}

This equation implies that the quantity inside the derivative is constant along the path of the guiding center, i.e., that
\begin{equation}
    \label{eqn:conserved quantity1}
    \frac{d\chi^\beta}{d\tau}\frac{d\chi_\beta}{d\tau} + 2\omega_0\rho_0^2\omega = -1.
\end{equation}
Using equation~(\ref{eq:rho_tau}) we can write this conserved quantity as
\begin{equation}
    \label{eqn:conserved quantity2}
    \frac{d\chi^\beta}{d\tau}\frac{d\chi_\beta}{d\tau} + 2\rho^2\omega^2 = -1.
\end{equation}

Note that we could have obtained the same result by taking directly the norm of the ansatz for the four-velocity, equation~\eqref{eqn:four-velocity}.

In the Newtonian limit, the second term in the above equation is just the magnitude of the gyrating velocity, which initially is the perpendicular velocity of the particle with respect to the magnetic field. The conservation law then simply states that the sum of the square of the norm of the guiding center velocity and the square of the gyrating velocity is equal to $-1$ one or, more precisely, that it converges to $-1$ to first order in gyroradius. Surprisingly, this is the same integral of motion found by \cite{Vandervoort_1960} in flat spacetime.

\section{Limiting Cases}
\label{sec: limiting cases}
In this section, we examine the behavior of the guiding center motion in various limiting cases in order to demonstrate explicitly that the covariant guiding center equations contain all drift velocities that were known previously. In particular, we consider: {\em (i)\/} a flat spacetime with a non-uniform electromagnetic field, to recover the $\Vec{E}\times\Vec{B}$ and $\vec{\nabla} B\times \vec{B}$ drifts and {\rm (ii)\/} a Schwarzschild spacetime to zeroth Newtonian order to recover the classical gravitational drifts. We then explore the equation of motion for the full Schwarzschild metric to analyze the drift velocities of particles in curved spacetime. Throughout this section, we reintroduce to the equations the various physical constants in order to compare the drift velocities in curved spacetime with the familiar drift velocities established in flat spacetime.

\subsection{Drift Velocities in Flat Spacetime}

We first consider the motion of a particle with mass $m$ and charge $q$ in a flat spacetime under the influence of an electric field $E_z$ in the $\hat{z}$-direction and a magnetic field $B_x$ in the $\hat{x}$-direction, with the latter varying as a function of $z$. We assume that the particle velocity $u_p^\beta\equiv dx^\beta/d\tau$ at proper time $\tau = 0$ is perpendicular to the direction of the magnetic field, i.e., that $u_{\rm p}^x=0$. 

In this case, the electromagnetic tensor is
\begin{equation}
F^{\alpha\beta}=\left[\begin{array}{cccc}
0 & 0 & 0 & -E_z\\
0 & 0 & 0 & 0\\
0 & 0 & 0 & -B_x\\
E_z & 0 & B_x & 0 
\end{array}\right]
\end{equation}
and the gyro-frequency $\omega_0$ is (see eq.~[\ref{eqn:eignevalues}])
\begin{align}
\label{eqn: omega flat cart}
    \omega_0 = \frac{q B_x}{m c}\left(1-\frac{E_z^2}{B_x^2}\right)^{1/2}\;,
\end{align}
which is the familiar result. To calculate the generalized radius of gyration $\rho_0$, we first evaluate the corresponding eigenvector of the electromagnetic field using equation~(\ref{eqn:Cayley-Hamilton Theorem}), i.e.,
\begin{gather}
\label{eqn: cart sigma}
\sigma^\alpha = \frac{1}{\sqrt{2}} \biggl(\frac{-i E_z/B_x}{
   \sqrt{1-E_z^2/B_x^2 }},0,\frac{-i}{
   \sqrt{1-E_z^2/B_x^2 }},1 \biggr),
\end{gather}
and then evaluate equation \eqref{eqn:gyroradius} to obtain
\begin{eqnarray}
\label{eqn: gyroradius flat cart}
    \rho_0 &=& \frac{mc}{\sqrt{2} q B_x(1-E_z^2/B_x^2)}\times \nonumber \\ 
    &&\qquad\left[ -\frac{E_z}{B_x} u_p^t  + u_p^y \right.\left.+i\left(1-\frac{E_z^2}{B_x^2}\right)^{1/2} u_p^z \right].
\end{eqnarray}

Denoting now the 4-velocity of the guiding center by $u^\beta$, we can use equation \eqref{eqn:GCA EOM Final} to write the guiding-center acceleration equations as
\begin{eqnarray}
\label{eqn: cart t original}
    \frac{d u^{t}}{d \tau} &=& \frac{q}{m\, c} \, E_z\,u^{z},\\
\label{eqn: cart x original}
    \frac{d u^{x}}{d \tau} &=& 0,\\
\label{eqn: cart y original}
    \frac{d u^{y}}{d \tau} &=& \frac{q}{c\, m} B_x\, u^{z},\\
    \frac{d u^{z}}{d \tau} &=& \frac{q}{m\, c}(E_z\,u^t -B_x\,u^y) \nonumber\\
    \label{eqn: cart z original}&&
    \qquad-\frac{q}{m\, c} \frac{\rho_0^2\, \omega_0\, }{\sqrt{1-E_z^2/B_x^2}} \frac{\partial B}{\partial z}\;.
\end{eqnarray}
Following the traditional approach, we assume that the acceleration terms on the left-hand-sides of these equations are of higher order and can, therefore, be neglected. Equations~(\ref{eqn: cart t original})-(\ref{eqn: cart y original}) immediately lead to $u^z=0$. Equation~(\ref{eqn: cart z original}) can then be solved for the only non-zero component of the drift velocity of the guiding center, $u^y$.

In order to recover the $\vec{E} \times \vec{B}$ drift velocity, we will first consider the case in which $\partial B/\partial z$ can also be neglected. Under these conditions, equation~\eqref{eqn: cart z original} simplifies to
\begin{align}
    \label{eqn: simplified z cart electric}
    0 = \frac{q}{m c}(E_z \, u^{t} - B_x\,u^{y}).
\end{align}
For non-relativistic particle velocities,  $u_{\rm p}^t=c$. Solving equation (\ref{eqn: simplified z cart electric}) for $u^y$ recovers the standard expression for the $\vec{E}\times\vec{B}$ drift in flat spacetime, i.e.,
\begin{align}
    \label{eqn: cart electric drift velocity}
    u^y = c \frac{E_z}{B_x} = \frac{c\vert\Vec{E}\times \Vec{B}\vert}{B^2}.
\end{align}

Returning to equation \eqref{eqn: cart z original}, we now reintroduce the terms that allow us to recover the $\vec{\nabla} B\times\vec{B}$ drift. As described in \cite{Vandervoort_1960}, this drift has a correction term for particles traveling at relativistic speeds under the influence of a perpendicular electric field. As a result, for this particular case, we assume that the 3-velocity of the particle speed may be comparable to the speed of light. In order to simplify the expression, we also assume that the particle velocity is perpendicular to both the electric and magnetic fields, i.e., that $u_p^x=u_p^y=0$.

As before, we examine equation~(\ref{eqn: cart z original}) after neglecting the acceleration term, i.e., we set
\begin{align}
    \label{eqn: steady state z cart magnetic}
    0= \frac{q}{m c} (E_z u^t - B_x u^y) - \frac{q}{m\, c} \frac{\rho_0^2\, \omega_0\,}{\sqrt{1-E_z^2/B_x^2}} \frac{\partial B}{\partial z}\;.
\end{align}

Using equations~\eqref{eqn: omega flat cart}, \eqref{eqn: gyroradius flat cart}, and \eqref{eqn: steady state z cart magnetic}, we can solve for the guiding center velocity in the $\hat{y}$-direction.
Note that all velocities in this expression are in terms of the proper time $\tau$. In order to put this expression in the traditional form, we write the particle 3-velocity perpendicular to the magnetic field as
\begin{equation}
    v_\perp= \frac{dz}{dt}=\frac{dz}{d\tau}\frac{d\tau}{dt}=u_p^z \left(\frac{u_p^t}{c}\right)=\gamma u_p^z\;,
\end{equation}
where $\gamma$ is the Lorentz factor corresponding to the particle velocity. Similarly, expressing the drift velocity in terms of coordinate time, we write $u^y=\gamma v^y$. The net result is
\begin{align}
   \label{eqn: grad B all terms}
    v^y = & - \frac{c E_z}{B_x} + \frac{\gamma m c v_\perp^2}{2q}  \frac{E_z^2}{B_x^4} \nabla B (1- \frac{E_z^2}{B_x^2})^{-2} \nonumber \\
    & + \frac{\gamma m c v_\perp^2}{2q} \frac{\nabla B}{2 B_x(1-\frac{E_z^2}{B_x^2})}\;.
\end{align}
The first term in the equation above is the $\vec{E} \times \vec{B}$ drift we calculated previously. The second and third terms describe the $\vec{\nabla} B\times\vec{B}$ drift velocity, along with their relativistic correction terms. Simplifying the second and third terms further, we recover the familiar form, i.e.,
\begin{align}
    v^y
    & = -c\frac{\vert\vec{E}\times\vec{B}\vert}{B^2} \nonumber\\
    &\qquad+\frac{\gamma m c u_\perp^2}{2 q B (1-\frac{E_z^2}{B_x^2})} \left\vert\vec{B} \times \nabla \biggl[ B \biggl(1-\frac{E_z^2}{B_x^2}\biggr) \biggr]\right\vert\;.
\end{align}
as also presented in \cite{Vandervoort_1960}.

\subsection{Newtonian Gravitational Drift}
\label{subsec:Newtonian Potential Case}

Next, we examine the drift introduced by a central gravitational field, in the Newtonian limit, in the presence of a constant magnetic field in the $\hat{\theta}$-direction. To simplify the algebra, we  assume that there is no electric field. To describe the gravitational field, we use the Schwarzschild metric in the zeroth Newtonian order
\begin{equation}
ds^2 = -(1-2GM/c^2r) dt^2 + dr^2 + r^2 (d\theta^2 + \sin{\theta}^2 d\phi^2)\;.
\end{equation}
We will also assume that the particle is non-relativistic and traveling on the equatorial plane at $\theta = \pi/2$.

The gyro-frequency of such a system (see eq.~[\ref{eqn:eignevalues}]) is similar to the one found in flat spacetime, but without the electric field contributions, i.e.,
\begin{align}
    \label{eqn: omega flat newt}
    \omega_0 = \frac{q B_\theta}{mc}\;.
\end{align}
The corresponding eigenvector is
\begin{gather}
\label{eqn: newtonian sigma}
\sigma^\alpha = \frac{1}{\sqrt{2}}\biggl(0,\,i,\,0,\,\frac{1}{r}\biggr),
\end{gather}
and the complex gyroradius is (see eq.~[\ref{eqn:gyroradius}])
\begin{align}
    \label{eqn: initial gyroradius newt}
    \rho_0 = \frac{m c}{\sqrt{2} q B_\theta} (- u_p^r+ i r u_p^\phi)\;.
\end{align}

Evaluating equation~\eqref{eqn:GCA EOM Final} in this configuration, we obtain the following set of equations for the acceleration of the guiding center:
\begin{eqnarray}
    \label{eqn: newtonian t original}
    \frac{d u^t}{d\tau} &=& \frac{-1}{(1- \frac{2 G M}{rc^2})} \frac{2 G M}{r^2 c^2} u^r u^t, \\
    \frac{d u^r}{d\tau} &=&- \frac{ q }{m\, c}B_\theta r\, u^\phi -\frac{G M}{c^2r^2} (u^t)^{2}  \nonumber \\ 
    \label{eqn: newtonian rr original}
    &&\qquad + r\,(u^\theta)^{2} + r \, (u^\phi)^{2}, \\
    \label{eqn: newtonian theta original}
    \frac{d u^\theta}{d\tau} &=& -\frac{2}{r} u^r u^\theta,\\
    \label{eqn: newtonian phi original}
    \frac{d u^\phi}{d\tau} &=& -\frac{2}{r} u^r u^\phi + \frac{q }{m \, c}\frac{B_\theta }{r} u^r.
\end{eqnarray}

As before, we neglect the acceleration terms such that equations~(\ref{eqn: newtonian t original})-(\ref{eqn: newtonian theta original}) result in $u^r=0$. We then set $u^t=c$ in equation~\eqref{eqn: newtonian rr original} and solve it for the {\em linear\/} velocity of the guiding center in the $\hat{\phi}$-direction to find
\begin{align}
    r u^\phi = -\frac{m\, c}{q B_\theta} \frac{G M}{r^2} = \frac{m\, c}{q} \frac{\vert\vec{g} \times \vec{B}\vert}{B^2}\;.
\end{align}
In the last term we have set $\vec{g}\equiv -(GM/r^2)\hat{r}$, recovering the standard expression for the gravitational drift velocity. 

\subsection{Gravitational Drift in the Schwarzschild Spacetime}

We now consider relativistic charged particle motion in the Schwarzschild spacetime 
\begin{eqnarray}
    ds^2 &=& -(1-2GM/rc^2) dt^2 + (1-2GM/rc^2)^{-1} dr^2 
    \nonumber\\
    &&\qquad+ r^2 d\theta^2 + r^2 \sin{\theta}^2 d\phi^2\;,
\end{eqnarray}
in the presence of a constant magnetic field in the $\hat{\theta}$-direction. As before, we  assume that there is no electric field.

The gyro-frequency for such a system is given by eq.~(\ref{eqn:eignevalues}), i.e.,
\begin{align}
    \label{eqn: omega schwartz}
    \omega_0 = \sqrt{1-\frac{2GM}{rc^2}} \frac{qB_\theta}{m c}.
\end{align}
This is analogous to the gyrofrequency in the Newtonian limit (cf.\ eq.~[\ref{eqn: omega flat newt}]) but with the addition of an associated redshift factor. Similarly, the associated eigenvector is
\begin{gather}
    \label{eqn: schwartz sigma}
    \sigma^\alpha = \frac{1}{\sqrt{2}}\Biggl(0,\, -i\sqrt{1-\frac{2GM}{c^2r}},\,0,\,\frac{1}{r}\Biggr)
\end{gather}
and the complex gyro-radius $\rho_0$ is (see eq.~[\ref{eqn:gyroradius}])
\begin{align}
    \label{eqn: qyroradius schwarz}
    \rho_0 = \frac{1}{\sqrt{1-\frac{2GM}{c^2r}}} \frac{mc}{\sqrt{2}qB_\theta}  \biggl( \frac{1}{\sqrt{1-\frac{2GM}{c^2r}}} u_p^r - ir u_p^\phi \biggr)\;.
\end{align}

The acceleration equation \eqref{eqn:GCA EOM Final} for this configuration becomes
\begin{align}
\label{eqn: schwarz r}
    \frac{du^r}{d\tau} = & 
    -\frac{q}{m\, c} B_\theta r\biggl(1-\frac{2GM}{c^2r}\biggr) u^\phi \nonumber \\
    & - \frac{GM}{c^2r^2} \biggl(1-\frac{2GM}{c^2 r}\biggr) (u^t)^2-\frac{G M}{c^2 r^2} \frac{1}{1-\frac{2GM}{c^2 r}} (u^r)^2\nonumber \\
    &\qquad\biggl(1-\frac{2GM}{c^2 r}\biggr) r (u^\theta)^2+\biggl(1-\frac{2GM}{c^2 r}\biggr) r (u^\phi)^2 \nonumber \\
    & \qquad \qquad - \frac{1}{2} \frac{G M}{c^2 r} \biggl[ \frac{(u_p^r)^2}{ (1-\frac{2GM}{rc^2})} + (r \,u_p^\phi)^2 \biggr]\;,
\end{align}
where, for brevity, we only display the important $r-$component. As in the Newtonian case, the other components of the acceleration equation lead to $u^r=0$.  Setting the radial acceleration also equal to zero and solving the resulting equation to first order in the {\em linear\/} velocity in the $\hat{\phi}$-direction, we obtain
\begin{align}
\label{eqn: u phi schwartz}
    r u^\phi = -\frac{m\, c}{q B_\theta} \frac{GM}{c^2 r^2} (u^t)^2 = \frac{m}{qc}\frac{\vert\vec{g} \times \vec{B}\vert}{B^2} (u^t)^2.
\end{align}
Here, we again defined $\vec{g}\equiv -(GM/r^2)\hat{r}$, for direct comparison to the Newtonian case, even though this is not the effective ``gravitational'' acceleration experienced by the particle in the Schwarzschild spacetime.

We now use the conservation law~\eqref{eqn:conserved quantity1} to write
\begin{eqnarray}
    u^\beta u_\beta + \biggl(-\frac{1}{\sqrt{1-\frac{2GM}{rc^2}}} u_p^r + ir u_p^\phi \biggr)^2 &=& -c^2\Rightarrow\nonumber\\
    u^tu_t + u^iu_j+\left(-\sqrt{g_{rr}} u_{rm p}^r+i\sqrt{g_{\phi\phi}}u^\phi_{\rm p}\right)^2&=&-c^2\;,
    \label{eqn: conservation schwartz}
\end{eqnarray}
where, in the last equation, we used $g_{rr}$ and $g_{\phi\phi}$ to denote explicitly some of the components of the Schwarzschild metric. Solving this equation for the time component of the guiding center velocity, we find
\begin{eqnarray}
    \label{eqn ut schwarzschild}
    u^t = \frac{\gamma c}{\sqrt{1-\frac{2GM}{c^2 r}}}\;,
\end{eqnarray}
where 
\begin{equation}
\label{eq:gamma schwarzschild}
    \gamma=\left(1-u^i u_i - g_{rr}u_p^r u_p^r-g_{\phi\phi}u_p^\phi u_p^\phi\right)^{1/2}\;.
\end{equation}
Here the $i-$index is running only over the spatial components of the 4-vectors. The 3-velocity of the guiding center $u^i$ will have a component along the magnetic field lines, which is equal to the same component of the particle velocity, i.e., $u^\theta=u^\theta_p$ in this case, and therefore, of zeroth order. It will also have a component perpendicular to the magnetic field lines, i.e., the drift velocity, which is of first order. Because we will be inserting the expression for $u^t$ into equation~(\ref{eqn: u phi schwartz}), which is already of first order, we keep only the zeroth order terms in equation~(\ref{eq:gamma schwarzschild}) to write
\begin{equation}
   \gamma=\left(1-g_{ij}u_p^i u_p^i\right)^{1/2}\;.
\end{equation}
This is simply the Lorentz factor associated with the particle velocity.

Substituting these into equation \eqref{eqn: u phi schwartz}, we obtain the gravitational drift velocity equation in the Schwarzschild spacetime
\begin{align}
    \label{eqn: gxb schwartz1}
    r u^\phi_{\vec{g} \times \vec{B}} = \frac{m\, c}{q} \frac{\gamma^2}{(1-\frac{2 G M}{c^2 r})} \frac{\vert\vec{g} \times \vec{B}\vert}{{B}^2}.
\end{align}

In order to convert this into an equation for the 3-velocity of the guiding center written in terms of coordinate time, we use equation~(\ref{eqn ut schwarzschild}) to obtain
\begin{align}
    \label{eqn: gxb schwartz2}
    r v^\phi_{\vec{g} \times \vec{B}} =r u^\phi_{\vec{g} \times \vec{B}} \left(\frac{u^t}{c}\right)
    =\frac{m\, c}{q} \frac{\gamma}{(1-\frac{2 G M}{c^2 r})^{1/2}} \frac{\vert\vec{g} \times \vec{B}\vert}{{B}^2}.
\end{align}

\section{Conclusions}
\label{sec:conclusion}

GRMHD has been the method of choice for simulating the global plasma dynamics around radiatively inefficient accreting black holes, despite the very large mean-free paths of particles in these systems. Kinetic approaches, on the other hand, are able to describe collisionless plasmas from first principles but are limited in the dynamic range they can simulate due to the large scale separation between the microscopic kinetic scales, i.e., the gyroradius and the plasma skin depth, and the macroscopic length scales of the systems. In this paper, we presented a set of covariant, guiding center equations of motion applicable to charged particles in general spacetimes and arbitrary electromagnetic fields. We derived conservation laws for the guiding center motion and showed that, in various limits, the equations simplify to the traditional expressions for the drift velocities.

Assuming that the electric field is sufficiently weaker than the magnetic field, we also derived the fully relativistic expressions for the various drift velocities in the Schwarzschild spacetime. For brevity, we present here only the final result
\begin{eqnarray}
    r u^\phi_{\vec{g} \times \vec{B}}
    &=&\frac{m\, c}{q} \frac{\gamma^2}{(1-\frac{2 G M}{c^2 r})} \frac{\vert\vec{g} \times \vec{B}\vert}{{B}^2},\\
    \label{eqn: dbxb schwartz}
    r u^\phi_{\nabla \vec{B} \times \vec{B}} &=& \frac{m\, c\, v_\perp^2}{2 q B} \frac{\gamma^2}{(1-\frac{2 G M}{c^2 r})} \frac{\vert\nabla B \times \vec{B}\vert}{B^2}, \\
    \label{eqn: exb schwartz}
    r u^\phi_{\vec{E} \times \vec{B}} &=&  \frac{\gamma}{(1-\frac{2 G M}{c^2 r})^{1/2}} \frac{c \vert\vec{E} \times \vec{B}\vert}{B^2}.
\end{eqnarray}
As expected, the drift velocities in the Schwarzschild spacetime follow similar relations with the counterparts in flat spacetimes, with the addition of the associated redshift and Lorentz factors. 

Finally, writing all equations for the drift velocities in terms of the corresponding 3-velocities using equation~(\ref{eqn ut schwarzschild}), we obtain
\begin{eqnarray}
    \label{eqn: gxb schwartz coord}
    r v^\phi_{\vec{g} \times \vec{B}} &=& \frac{m\, c}{q} \frac{\gamma}{(1-\frac{2 G M}{c^2 r})^{1/2}} \frac{\vert\vec{g} \times \vec{B}\vert}{{B}^2}, \\
    \label{eqn: dbxb schwartz coord}
    r v^\phi_{\nabla \vec{B} \times \vec{B}} &=& \frac{m\, c\, v_\perp^2}{2 q B} \frac{\gamma}{(1-\frac{2 G M}{c^2 r})^{1/2}} \frac{\vert\nabla B \times \vec{B}\vert}{B^2}, \\
    \label{eqn: exb schwartz coord}
    r v^\phi_{\vec{E} \times \vec{B}} &=&   \frac{c \vert\vec{E} \times \vec{B}\vert}{B^2}.
\end{eqnarray}
It is interesting to note that the  redshift factor appears in the gravitational drift and in the $\nabla B\times \vec{B}$ drift but not in the $\vec{E}\times\vec{B}$ drift. In the first case, the redshift factor corrects the Newtonian gravitational acceleration so that the latter becomes infinite at the horizon. In the second case, the redshift factor corrects the spatial derivative of the magnetic field strength, which needs to be expressed in terms of proper and not coordinate distances. Since the $\vec{E}\times\vec{B}$ drift depends on neither the gravitational acceleration nor on derivatives, it does not contain any redshift corrections.

\begin{acknowledgements}
This work has been supported by NSF PIRE award OISE-1743747. T.T. acknowledges support from the Alfred P. Sloan Foundation and the Ford Foundation.
\end{acknowledgements}

\bibliography{main}
\end{document}